\documentstyle[prl,aps]{revtex}
\begin{document}
\title{Critical aspects of hierarchical protein folding}
\author{Alex Hansen, \footnote{Permanent address: Department of Physics,
Norwegian University of Science and Technology, NTNU, N--7034 Trondheim, 
Norway}\footnote{Electronic Address: Alex.Hansen@phys.ntnu.no}
Mogens H. Jensen,\footnote{Electronic Address: mhjensen@nbi.dk}
Kim Sneppen\footnote{Electronic Address: sneppen@nbi.dk} 
and Giovanni Zocchi\footnote{Electronic Address: zocchi@nbi.dk}}
\address{Niels Bohr Institute and NORDITA, Blegdamsvej 17, DK-2100 {\O}, 
Denmark}
\date{\today}
\maketitle
\begin{abstract} 
We argue that the first order folding transitions of proteins
observed at physiological chemical conditions end in a critical
point for a given temperature and chemical potential
of the surrounding water. We investigate this critical point
using a hierarchical Hamiltonian and determine its universality class.
This class differs qualitatively from those of other known models.
\end{abstract} 
\vskip0.5cm

Proteins are complex macromolecular objects \cite{ablrrw94} whose structure
is determined by the evolutionary process that formed them \cite{n84,g85}.
In particular this means that the folding of proteins into its
three-dimensional structure must be efficient. 
There have been several attempts to grasp aspects of the protein folding
process \cite{dbyfytc95}, in enumeration of configurations \cite{lb97}, in
description of folding pathways \cite{dfc93}
and in discussing the influence of water on protein 
structure \cite{ltw97}.

Presumable proteins evolved by subsequently adding properties to simpler 
structures which also fold. This has led us to propose
a hierarchical description of the folding process,
where each added subunit folds conditionally
on the folding of structures above it in the hierarchy \cite{hjsz98a}.
Formulating this in the framework of statistical mechanics 
we suggest that the folding can be parametrized
in a number of variables $\varphi_1,..., \varphi_N$ each taking 
the value $0$ or $1$ where $1$ correspond to a correctly folded 
substructure \cite{hjsz98a,hjsz98b}. The hierarchy is implemented 
through the Hamiltonian
\begin{equation}
\label{eq1}
H~=~-\ {\cal E}_0\ (\varphi_1+\varphi_1\varphi_2+
 \varphi_1\varphi_2\varphi_3+\cdots +
 \varphi_1\varphi_2\cdots\varphi_N)~~~.
\end{equation}
The interactions between the protein and the surrounding water may be
taken into account by adding
to this Hamiltonian a coupling parametrized through the water variables 
$w_1, w_2, ... , w_N$ \cite{hjsz98b}.  These variables couple to
the unfolded protein degrees of freedom 
because they expose hydrophobic amino acids to the water.
The resulting Hamiltonian is 
\begin{equation}
\label{eq2}
H~=~- {\cal E}_0 ~
(\varphi_1+\varphi_1\varphi_2+
\varphi_1\varphi_2\varphi_3+\cdots +
\varphi_1\varphi_2\cdots\varphi_N)\\ 
- [ (1-\varphi_1) w_1 + (1-\varphi_1\varphi_2) w_2 + ... + 
(1-\varphi_1\varphi_2\cdots\varphi_N) w_N ]   
\end{equation}
where the hydrophobic effect is taken into account by 
letting each of the $w$'s take a value from the set, 
${\cal E}_{\min}+s\Delta$, $s=0,1,...,g-1$. $\Delta$ is
the spacing of the energy levels of the water-protein interactions.

The partition function is
\begin{equation}
\label{part}
Z \;=\;  
\left( e^{{\cal E}_0 /T} \right)^{N} ~  
\left( \frac{1}{2} ~ \frac{r^{-N}-1}{1-r} ~~+~ 1 \right)
\end{equation}
The variable $r$ is the ratio of statistical weights of unfolded
to folded state, per variable:
\begin{equation}
\label{rrr}
r ~=~ \frac{g}{2}~ e^{-\mu/T}~  
\frac{1-e^{-\Delta/T}} 
{1-e^{- g \Delta/T}} 
\end{equation}
with $\mu=-{\cal E}_0-{\cal E}_{\min}$ being 
the chemical potential of the surrounding water.

The physical meaning of this model is that the water 
molecules in contact with an unfolded portion of the protein 
has lower entropy than when not in contact (thus in the our model,
hydrophobicity is caused by ordering of water and not by
repulsive potentials, as is usually believed \cite{e35}). 
In the model one finds that a first order transition  
takes place when the parameter $r$ switches between $r<1$ and $r>1$.
Plotting $r$ against $T$ one obtains a non-monotonic 
function which for small $\mu$ values 
passes $r=1$ twice, corresponding to unfolding at both low
and high temperature, as indeed seen in experiments \cite{p90,p95}.
The mechanism for the transitions is the following.  
At high temperature the entropy gain of the protein chain
causes the unfolding. As temperature is lowered the system 
gains more entropy by shielding the hydrophobic residues 
from the water.  This leads to folding.
As the temperature is lowered even further
the cold unfolding transition occurs. Below this transition
entropy is insignificant and the dominating effect is the 
attractive coupling between the water and the unfolded protein.

For an intermediate value of the chemical potential,
$r$ just touches the line $r=1$, that is $dr/dT=0$ when $r=1$,
corresponding to a merging of two first order transitions.
This defines a critical point.  Around this point, $r$ varies quadratically in
$T-T_c$ and linearly in $\mu-\mu_c$, as seen from expanding Eq.\ (\ref{rrr}).
In experiments of protein folding this point is 
accessible by changing the pH value of the solution.
In fact, Privalov's data on low pH values indeed 
indicate that such a critical point exists.
The scaling properties around this point
thus opens for a possibility to gain insight into the 
nature of the folding process, in particular whether 
the hierarchical scheme we suggest can be falsified.

In Fig.\ \ref{fig1}a we show heat capacity as a function of temperature
for chemical potential below, at and above the critical value
$\mu=\mu_c$. 
For the chosen values of ${\cal E}_0=1$ and level density
$\Delta=0.02$ and $g=350$
the critical point is situated at $T_c=1.33303\dots$, $\mu_c=1.2838\dots$.
That is, it is situated at a {\it minimum\/} of the heat capacity curve.
This is at first sight surprising, usually heat capacity has 
a pronounced increase at the critical point.
The minimum reflects a partial ordering of the hierarchy, 
as envisioned in Fig.\ \ref{fig1}b where we show the degree 
of folding, counted by the average number of folded 
variables $\varphi_i=1$, $i=1,...,n$ from $i=1$ 
until the first variable $i=n+1$ which takes value $\varphi_{n+1}=0$.
The average value of this $\langle n \rangle$ is
$N/2$ at the critical point, reflecting 
that the system is on average half ordered at this point.
Correspondingly the heat capacity dips to a value
in between the value of an unfolded and a
completely folded state.

To characterize the functional form of the dip
in the heat capacity, we investigate analytically
$C_{sing}(T)=C(T,\mu)-C(T,\mu_c)$ with $\mu>>\mu_c$
for different values of hierarchy size $N$.
For finite $N$ we may express the singular
part of the heat capacity in the form:
\begin{equation}
C_{sing} ~ =~ |T_c-T|^{-\alpha} ~ g \left( (T_c-T) N^{1/\nu} \right)
\end{equation}
where $g(x) \rightarrow const$ when $x\rightarrow \infty$
and $g(x)\propto x^{\alpha}$ when $x\rightarrow 0$.
We find analytically $\alpha=\nu=2$ from differentiating 
the partition function (\ref{part}).
Fig.\ \ref{fig2}a demonstrate this finite size scaling.
Similarly we in Fig.\ \ref{fig2}b show the behavior
of the order parameter $\langle n \rangle$
as function of $T-T_c$ and $N$:
\begin{equation}
\langle n \rangle ~ =~ |T-T_c|^{\beta} ~ 
f \left( (T-T_c) N^{1/\nu} \right) 
\end{equation}
with $f(x) \rightarrow const$ when $x\rightarrow \infty$
and $f(x)\propto x^{-\beta}$ when $x\rightarrow 0$
where exponents $\beta=-2$, also found analytically.  
It may be surprising that $\beta$ 
is negative, but this reflect in part
the unusual use of an extensive (in $N$) order parameter,
in part that for $\mu=\mu_c$ then the order parameter only obtains 
a non-zero value at $T=T_c$ when $N\rightarrow \infty$.

Likewise, we find that the susceptibility 
$\chi=d\langle n\rangle/d\mu$ scales as 
$|T-T_c|^{-\gamma}$ where $\gamma=4$ and 
that $\langle n \rangle \propto (\mu-\mu_c)^{1/\delta}$ 
for $\mu>\mu_c$ where $\delta=-1$. Thus the usual exponent relations,
$\alpha+2 \beta + \gamma=2$, $\alpha+\beta (\delta+1) =2$,
and $\gamma (\delta + 1) =(2-\alpha)(\delta-1)$ 
are fulfilled \cite{s71}.
However the hyperscaling reation $d \nu = 2-\alpha$, where $d$ is the
dimensionality of the system, is not fulfilled.  However, this relation has
no meaning, as there are no spatial degrees of freedom.

In terms of experiments on proteins, the relevant
scaling behaviour is the how the degree of folding
(order parameter) and the heat capacity behaves
as function of temperature, when one changes chemical potential
away from its critical value.
The qualitative prediction is that the width 
of the singular part of the heat capacity has a minimum
at the critical value $\mu=\mu_c$.
The broadening of the heat capacity is
\begin{equation}
C_{sing} (T-T_c)^2 ~=~ 
h\left(\frac{T-T_c}{\Delta \mu^{1/2}}\right) ~~ for~~ 
\mu> \mu_c~~ 
\end{equation}
where $h(x)\propto x^{-2}$ for $x \rightarrow \infty$ and 
$h(x)=const$ for $x \rightarrow 0$
and where $\Delta \mu=\max(\mu-\mu_c, \Delta \mu_{\min})$
with $\Delta \mu_{\min} \propto 1/N$ takes into account the
finite size sensitivity of the scaling. We show in Fig.\ \ref{fig3}a,
an example of such a data collapse.
These predictions are experimentally accessible through the use of standard
calorimetric techniques, where one should seek to 
obtain a data collapse above the critical point, 
i.e. the point of minimal width.
The heat capacity below the critical $\mu$ is complicated by
the merging of two first order transitions.
However, the distance between these moves away from each other
in $T$ as $\Delta \mu^{1/2}$.

Likewise, we expect the degree of folding $\langle n \rangle$
to show data collapse of the form
\begin{equation}
\langle n \rangle (T-T_c)^2
~=~ k\left(\frac{T-T_c}{\Delta \mu^{1/2}}\right) ~~ for~~ 
\mu> \mu_c~~ 
\end{equation}
where $k(x)$ behaves asymptotically as $h$. We show this in
Fig.\ \ref{fig3}b.
This quantity can be observed experimentally through fluoresence measurements.

In summary, we have proposed that the hierarchical ordering 
of folding pathways implies a critical point with 
a diminishing heat capacity at criticality.
We have determined all critical exponents, and proposed two
experiments that could confirm or falsify the concept
of hierarchical folding.

\begin{figure}
\caption{a) Heat capacity, $C$, as a function of $T$.
b) Degree of folding, $\langle n\rangle$, as a function of $T$.  
Here $g=350$, $\Delta=0.02$ and $N=100$. The value $N=100$ has been chosen
as to be close to realistic values for this parameter.
\label{fig1}
}
\end{figure}
\begin{figure}
\caption{a) Finite size scaling of the heat capacity for $\mu=\mu_c$,
$g=350$ and $\Delta=0.02$. Here $\alpha=2$ and $\nu=2$.
b) Finite size scaling of degree of folding, $\langle n\rangle$.  Here
$\beta=-2$.
\label{fig2}
}
\end{figure}
\begin{figure}
\caption{a) $C_{sing}(T-T_c)^2$ {\it vs.\/} $(T-T_c)/\Delta\mu^{1/2}$.
b) $\langle n\rangle(T-T_c)^2$ {\it vs.\/} $(T-T_c)/\Delta\mu^{1/2}$.
We have chosen $N=100$, $g=350$ and $\Delta=0.02$.  Note the good quality
of the data collapse in spite of smallness of the system.
\label{fig3}
}
\end{figure}
\end{document}